\documentclass[runningheads]{llncs}
\usepackage{booktabs} % For formal tables
\usepackage{amssymb}

\usepackage{amsthm}
\usepackage{afterpage}
\usepackage{mathtools}
\usepackage{graphicx}

\usepackage[ruled,vlined,linesnumbered]{algorithm2e}
\usepackage{url}
\usepackage{float}%figure environment
\usepackage{booktabs}
\usepackage{todonotes}

\usepackage{mdframed}
\usepackage{wrapfig}
\usepackage{rotating}
\usepackage{multirow}
\usepackage{multicol}
\usepackage{tabularx}
\usepackage{longtable}
\usepackage{comment}
\usepackage{caption}
\usepackage{pifont}
\usepackage{siunitx}
\usepackage[T1]{fontenc}
\usepackage{adjustbox}
\usepackage{color}

\usepackage{graphicx}
\usepackage{subfigure}
\usepackage{url,hyperref}
\usepackage{xspace}

\usepackage{siunitx}
\usepackage{array}
\usepackage{todonotes}
\usepackage{placeins}
\usepackage{afterpage}
\usepackage{lscape}

\usepackage{mathtools}
\usepackage{commath}

\let\norm\undefined % <-- "Undefine" \norm
\DeclarePairedDelimiter\norm{\lVert}{\rVert}

    \newcommand{\naivecol}{\textsc{NaiveCol}\xspace}
     \newcommand{\grecond}{\textsc{GreConD}\xspace}
     \newcommand{\hyper}{\textsc{Hyper}\xspace}
     \newcommand{\greess}{\textsc{GreEss}\xspace}
     \newcommand{\panda}{\textsc{PaNDa}\xspace}
     \newcommand{\grecondplus}{\textsc{GreConD+}\xspace}

 \usepackage{color}
\definecolor{pblue}{rgb}{0.13,0.13,1}
\definecolor{pgreen}{rgb}{0,0.5,0}
\definecolor{pred}{rgb}{0.9,0,0}
\definecolor{pgrey}{rgb}{0.46,0.45,0.48}
\definecolor{DarkGreen}{rgb}{0,0.95,0}

\usepackage{listings} 
\lstdefinestyle{sparql}{float=tb, numberblanklines=true, morekeywords={SELECT,DISTINCT,SAMPLE,FROM,WHERE,FILTER,ORDER,GROUP,BY,IN,AS,GRAPH,SERVICE,PREFIX}}

\tikzset{filled/.style={fill=circle area, draw=circle edge, thick},
    outline/.style={draw=circle edge, thick}}

\begin{document}
% \begin{itemize}
%     \item Introduction \\
%           AFP \& DBP  problem
% \item methods:
% \begin{itemize}
%     \item 1. GreCond(Algorithm 2)
%     \item 2. Asso Ok
    
%     \item 3. GreCond+
%     \item 4. NaivCol  similar but cols only and no ext.
%     \item 5. 8M
% \end{itemize}
% \item  datasets ok
%   Mushroom, Chess, Palea, DBLP, Firewall 1,
%   Emea,tic\_tac\_toe,customer
% 4.Related work\\
%  5.Metrics: coverage, minK\\
%          plot coverage with k
%          running time plot SWDF(

% \end{itemize}
%     \todo{describe algorithm stage1.
%     pseudocode stage2
%     run experiments: DBP, AFP}

\title{topFiberM:  Scalable and Efficient Boolean Matrix Factorization}
\author{Abdelmoneim Amer Desouki\inst{1}
% \orcidID{0000-0003-2083-1277} 
\and Michael R\"oder\inst{1}
\and Axel-Cyrille Ngonga Ngomo\inst{1} }

\authorrunning{A. Desouki et al.} % abbreviated author list (for running head)
\institute{
Data Science Group, Paderborn University
%   \city{Paderborn}
%   \state{Germany}
\\Warburgerstrasse 100, 33098 Paderborn\\
	\email{desouki@mail.upb.de, michael.roeder@upb.de, axel.ngonga@upb.de}
}

\maketitle              % typeset the header of the contribution
\begin{abstract}
Matrix Factorization has many applications such as clustering. 
When the matrix is Boolean it is favorable to have Boolean factors too. 
This will save the efforts of quantizing the reconstructed data back, which usually is done using arbitrary thresholds. 
Here we introduce topFiberM a Boolean matrix factorization algorithm. 
topFiberM chooses in a greedy way the fibers (rows or columns) to represent the entire matrix. 
Fibers are extended to rectangles according to a threshold on precision. 
The search for these "top fibers" can continue beyond the required rank and according to an optional parameter that defines the limit 
for this search. 
A factor with a better gain replaces the factor with minimum gain in "top fibers". 
We compared topFiberM to the state-of-the-art methods, it achieved better quality for the set of datasets usually used in literature. 
We also applied our algorithm to linked-data to show its scalability. 
topFiberM was in average 128 times faster than the well known \texttt{Asso} method when applied to a set of matrices representing
 a real multigraph although \texttt{Asso} is implemented in C and topFiberM is implemented in R which is generally slower than C. 
 topFiberM is publicly available from Github (https://github.com/dice-group/BMF).

\end{abstract}

% ---------------------------------------

\section{Introduction} 
 
Boolean Matrix Fatorization (BMF) for a Boolean Matrix $I (n \times m)$ tries to find two Boolean matrices $A (n \times k)$ and $B (m \times k)$ such that $I\approx (A \circ B)$ where $(A \circ B)_{ij} = max^{k}_{l=1} min(A_{il} , B_{lj} )$ and k is the rank of the factorization. There are two types of problems in Boolean Matrix factorization. The Discrete Base Problem (DBP)\cite{miettinen2008discrete} and Approximate Factorization Problem \cite{belohlavek2010discovery}. In the first problem it is important to find the best coverage over small rank values while in the second problem it is required to find the minimum rank to get error less than a defined threshold\cite{belohlavek2018toward}
\subsection{Related Work}
We introduce some of BMF methods:

\emph{Asso}~\cite{miettinen2008discrete}: is the most famous BMF method. It constructs a square matrix (i.e. Association matrix) with the same number of columns  as the data matrix and employs a three parameters to form the desired factor matrices in a greedy way. It is implemented in C by the authors.

\grecond~\cite{belohlavek2010discovery} uses the Formal Concept Analysis (FCA)\cite{ganter2012formal} to compute exact decompositions. The algorithm can be used to provide approximate DBP solutions by stopping it after computing the first k factors. Also it can provide approximate solution to AFP by stopping it after the committed error E does not exceed $\epsilon$~\cite{belohlavek2018toward}. 

\grecondplus~\cite{belohlavek2010discovery} is an extension to \grecond that can backwardly change factors when new better ones are discovered. 
The results demonstrate that it outperforms other methods in terms of quality and robustness to small changes in data but this came on a price of being much slower.

\naivecol\cite{ene2008fast} is a simple algorithm which searches for factors in a greedy way in the columns of the input matrix I. The algorithm may be used for both AFP and DBP\cite{belohlavek2018toward}.
% \naivecol is very fast.

For methods \hyper~\cite{xiang2011summarizing}, \greess~\cite{belohlavek2015below} and \panda~\cite{lucchese2010mining}, we refer to a recent review of BMF methods in \cite{belohlavek2018toward}. 

Metrics used to evaluate BMF methods mainly include coverage which include the two types of errors and is defined as follows:
\begin{equation}
   c = 1 - \norm{I -(A \circ B)}/\norm{I}
   \end{equation}

We introduce topFiberM a BMF method which can be seen as an extension to \naivecol by including not only columns but also rows and by extending the coverage of factors.
topFiberM also includes parameters like \emph{Asso} that parameters can be used to trade between recall and precision of results. Moreover fibers (rows/columns) are rectangles with one side equal to unity.
topFiberM keeps track of uncovered ones in I and can eliminate factors when better ones are discovered. topFiber rely on matrix operations and not single bit-operations which makes it very fast when compared to other methods.
% topFiberM is very fast as it uses floating po\texttt{in}t operations in $\big{O}(m k)$.

\section{topFiberM}
The algorithm starts by initializin\texttt{g }the factor search matrices As and Bs to zeros, \texttt{i} to 1, input matrix $I$ is copied to \texttt{X} and \texttt{SR} parameter is set to not exceed the minimum dimension. 
% In each iterationor n and it X represents the number of factors to search in. Then it itertFinally the A and B matrix are taken from the As and Bs according to the indexes in \texttt{tf}. il all ones in I are covered.
In each iteration the row sum and column sum is calculated for the not yet covered ones in \texttt{X}. The fiber with the maximum number of ones is considered.
If it is a row ( value in \texttt{mxr}) then we try to extend this by finding rows in \texttt{X} which are similar to mxr according to parameter \texttt{tP}. 
The algorithm accepts first \texttt{k} fibers and stores information about them in \texttt{tf} including index and gain. Note that when calculating the gain the true positives are calculated depending on \texttt{X} (not yet covered) while the false positives are calculated on the original matrix $I$. 
Afterwards the ones that are covered in this iteration are set to zeros to exclude from further processing. We used R function \texttt{expand.grid} to get the result of applying one factor. This function finds the permutations of two vectors which are in this case the indexes of ones in factors. The true positives will be the sum of values in $I$ of the resulting set of indexes ( variable \texttt{ix}) and the false positives will be $sum(!I[ix])$.
When \texttt{i} is greater than \texttt{k} we compare the gain to the minimum gain in \texttt{tf} (top fibers) if the gain is not more than the minimum the fiber is marked as excluded. Otherwise we replace that fiber with minimum gain in the \texttt{tf} list and recalculate the uncovered after considering the new fiber.
In each iteration the matrix \texttt{X} is checked if it is all zeros then the algorithm terminates. 
Finally the A and B matrix are taken from the As and Bs according to the indexes in \texttt{tf}. See Algorithm~\ref{alg:topFbr}.

\begin{algorithm}[!htbp]
     \SetAlgoLined
     \KwData{I: $n \times m$ Boolean matrix, k: rank,
     tP: threshold on precision,
     SR: search limit of factors default is k}
     \KwResult{Boolean factors A: $n \times k$, B: $m \times k$ where $X\approx A \circ B$}
     
    SR = min(SR, m, n)\;
    As = zeros(n, SR)\;
	Bs = zeros(SR, m)\;
	tf = array(struct(i, fiber type, fiber index, gain))\;
	excluded\_cols = rep(FALSE, m)\;
    excluded\_rows = rep(FALSE, n)\;
	X = I\;
	i=1\;
	\While{($i \leq SR$ )}{
        Find rowSums and columnSums of X\;
        Choose best Fiber (has maximum no of ones)\;
    	let best unexcluded fiber be a row with index mxr\;
          Bi = X[mxr, ] \;
          rtp = rowSums(X[, Bi])\;
          //better to be calculated on original I
    	  rfp = rowSums(!I[, Bi])\;
		  Ai = $ifelse(rtp = 0, FALSE, ((rtp/(rtp+rfp)) \geq tP))$\;
          //revise B
          ctp = colSums(X[Ai,])\;
		  cfp = colSums(!I[Ai,,drop=FALSE])\;
		  Bi = ifelse(ctp=0,FALSE,((ctp/(ctp+cfp)) >=tP))\;
    	  ix = as.matrix(expand.grid(which(Ai),which(Bi)),ncol=2)\;
		  gain = sum(X[ix]) - sum(!I[ix])\;
         Similarly when the best fiber is a column but with reverse role of Ai and Bi.\\
        %  }
            As[,i] = Ai\;
            Bs[i,] = Bi\;
        \eIf{$i \leq k$}{
          X[ix] = FALSE\;
          add row to tf with i, fiber type ( 1 for a row and 2 for a column), mxr and gain\\
        }{
           \eIf{if gain $\leq$ min gain in tf}{
                    mark fiber as excluded
                }{
                    replace row with min gain in tf with current fiber\\
                    X = uncovered ones in I\\
                }
        } 
         // test if all ones are covered\\
         \If{sum(X) = 0} { 
                   break\;
         }
        i = $i + 1$\;
    }%while
    A = columns of As with index in tf\;
    B = rows of Bs with index in tf\;
\caption{topFiberM: with backward correction}
\label{alg:topFbr}    
\end{algorithm}
 
\section{Evaluating topFiberM}
\subsection{Applying to literature datasets}
We followed similar approach as in~\cite{belohlavek2018toward} to evaluate the performance of our algorithm. The datasets used are Chess~\cite{bache2013uci}, DBLP~\cite{miettinen2009matrix}, Firewall 1~\cite{ene2008fast}, Mushroom~\cite{bache2013uci}, and Paleo\footnote{NOW public release 030717, available from http://www.helsinki.fi/science/now/}\cite{Fortelius2003Now}. The Boolean matrices was downloaded as MATLAB mat files from http://datasets.inf.upol.cz/. The main properties of the datasets are shown in Table~\ref{tab:ds}.

\begin{table}[!htb]
    \centering
    % \sisetup{
    %     table-column-width      = 1cm ,
    %     table-number-alignment  = right,
    %     group-separator         = {,},
    %     group-four-digits       = true
    %     }
    \begin{tabularx}{\columnwidth}{@{}p{2.5cm}XXXX}
    % \begin{tabularx}{\columnwidth}{@{}SSSSSS}
         \toprule
         {Dataset}	 &{Dimensions}	&{Number of 1s} &{Density}  \\
         \midrule
         Chess	  &$3196 \times 76$   &118 252	  &0.487\\
         DBLP     & $6980 \times 19$	& ~17 173 &0.130\\
         Firewall 1	&$365 \times 709$	& ~31 951 &0.124\\
         Mushroom & $8124 \times 119$	&186 852  &0.193 \\
         Paleo	  &$501 \times 139$&	~ ~3 537  &0.051\\
         \bottomrule
    \end{tabularx}
    \caption{Datasets used}
    \label{tab:ds}
\end{table}
\subsection{DBP view}
Table~\ref{tab:cv} represents the results of topFiberM compared to three BMF methods that performed best in the evaluation of BMF methods in \cite{belohlavek2018toward} (\grecondplus outperforms \grecond). 
Out of 20 values topFiberM achieved the best coverage in 16 and the well-known BMF method \emph{Asso} achieved best in 5 times.  
% In case of Mushroom dataset topFiberM take a row with all ones and repeated it 13 times for other 12 similar rows while \grecond calculated a rectangle of 1728 by 14.
% From the above results although topFiberM can be used for DBP it may not give good results at very small ranks.
The parameter used for tP is 0.5, this means to minimize the overall error and equally cost of the two types of error. The parameter \texttt{SR} value was set to 100.  
From these results it appears that topFiberM outperforms other methods in DBP. 
\begin{table*}[!htbp]
    \centering \footnotesize
\begin{tabularx}{\columnwidth}{@{}XXXXXX}
          \toprule
Dataset&	k	&Asso	&\grecondplus &NaiveCol	&topFiberM\\

\midrule
Chess & 1 & 0.497 & 0.447 & 0.119 & \textbf{0.610}\\
      & 2 & 0.574 & 0.506 & 0.177 & \textbf{0.625}\\
      & 5 & 0.628 & 0.621 & 0.323 & \textbf{0.667}\\
      & 10 & 0.703 & 0.710 & 0.461 & \textbf{0.724}\\
      
DBLP & 1 & 0.111 & 0.131 & 0.111 & \textbf{0.187}\\
     & 2 & 0.217 & 0.238 & 0.217 & \textbf{0.293}\\
     & 5 & 0.475 & 0.468 & 0.475 & \textbf{0.495}\\
     & 10 & \textbf{0.738} & 0.692 & \textbf{0.738} & \textbf{0.738}\\

Firewall 1 & 1 & \textbf{0.726} & 0.688 & 0.651 & 0.724\\
           & 2 & 0.818 & 0.841 & 0.804 & \textbf{0.847}\\
           & 5 & 0.908 & 0.951 & 0.932 & \textbf{0.953}\\
           & 10 & 0.917 & 0.979 & 0.976 & \textbf{0.980}\\
           
Mushroom & 1 & 0.226 & 0.131 & 0.129 & \textbf{0.253}\\
         & 2 & \textbf{0.323} & 0.235 & 0.234 & 0.305\\
         & 5 & 0.461 & \textbf{0.504} & 0.398 & 0.425\\
        & 10 & 0.555 & \textbf{0.613} & 0.512 & 0.522\\
 
Paleo & 1 & \textbf{0.027} & \textbf{0.027} & \textbf{0.027} & \textbf{0.027}\\
      & 2 & 0.047 & 0.047 & 0.047 &\textbf{0.049}\\
      & 5 & 0.105 & \textbf{0.106} & 0.105 & \textbf{0.106}\\
      & 10 & \textbf{0.182} & 0.181 & \textbf{0.182} & \textbf{0.182}\\
\midrule
Count best &  & 5 & 4 & 3 & \textbf{16}\\
         \bottomrule
    \end{tabularx}
    % \end{tabular}
    \caption{Coverage of low ranks (DBP view), bold values are the best in row. Results of \texttt{Asso} and \naivecol were taken from ~\cite{belohlavek2018toward}}
    \label{tab:cv}
\end{table*}

\subsection{AFP view}
Table~\ref{tab:afp} represents the results of topFiberM compared to other three BMF methods for AFP view as in the recent evaluation of BMF methods in \cite{belohlavek2018toward}. Out of 20 values topFiberM achieved the best coverage in 17 then \naivecol got 8, \emph{Asso} was not able to run for high ranks of bigger datasets (scalability problem) while \grecondplus achieved the best in 5 values. From these results we can conclude that topFiberM can be used to solve AFP. 
The parameter used for \texttt{SR} is k+10, for tP values are higher as shown in Table~\ref{tab:afp} column tP. As can be seen from results the value of tP that gives best coverage increases with the coverage (c).
The parameter w in \grecondplus was set to 4 as suggested by \cite{belohlavek2018new}. 
% tried with values 0.5, 1, 4, and 8 and we took the best result.
\begin{table*}
    \centering \footnotesize
\begin{tabularx}{\columnwidth}{@{}p{2cm}p{1cm}p{1cm}XXXp{0.5cm}@{}}
          \toprule
\textbf{Dataset}&	\textbf{\emph{c}}	&\textbf{Asso}	&\textbf{\grecondplus}	&\textbf{\naivecol}	&\textbf{topFiberM} &tP\\

\midrule
Chess & 0.8  & 21 & \textbf{19} & 21 & \textbf{19} & 0.7\\
      & 0.9  & NA & 34 & 34 & \textbf{33} & 0.8\\
      & 0.95 & NA & 48 & 46 & \textbf{45} & 0.8\\
      & 1    & NA & 130 & 72 & \textbf{71} & 1\\
DBLP & 0.8 & \textbf{12} & 13 & \textbf{12} & \textbf{12} & 0.7\\
     & 0.9 & \textbf{15} & 16 & \textbf{15} & \textbf{15} & 0.7\\
     & 0.95 & \textbf{17} & 18 & \textbf{17} & \textbf{17} & 0.7\\
     & 1 & \textbf{19} & 21 & \textbf{19} & \textbf{19} & 0.7\\
Firewall 1 & 0.8  & \textbf{2} & \textbf{2} & \textbf{2} & \textbf{2} & 0.7\\
           & 0.9  & 5 & 4 & 4 & \textbf{3} & 0.7\\
           & 0.95 & NA & 6 & 7 & \textbf{5} & 0.7\\
           & 1    & NA & 100 & 71 & \textbf{69} & 0.9\\
Mushroom & 0.8  & 47 & \textbf{29} & 32 & 34 & 0.8\\
         & 0.9  & NA & \textbf{46} & 47 & 50 & 0.9\\
         & 0.95 & NA & \textbf{62} & \textbf{62} & 65 & 0.9\\
         & 1    & NA & 120 & 110 & \textbf{109} & 0.9\\
Paleo & 0.8 & 84 & 86 & \textbf{83} & \textbf{83} & 0.7\\
      & 0.9 & 109 & 110 & 107 & \textbf{106} & 0.7\\
      & 0.95 & 125 & 127 & 122 & \textbf{121} & 0.7\\
      & 1 & NA & 151 & \textbf{139} & \textbf{139} & 0.8\\
 \midrule
 Count best &  & 5 & 5 & 8 & \textbf{17}\\
         \bottomrule
    \end{tabularx}
    % \end{tabular}
    \caption{Minimum ranks to get coverage (AFP view), bold values are the best in row. Results of \texttt{Asso} and \naivecol were taken from ~\cite{belohlavek2018toward}}
    \label{tab:afp}
\end{table*}

\subsection{Applying to RDF data}
 Resource Description Framework (RDF)\footnote{\url{https://www.w3.org/RDF/}} is used to represent data in form of triples i.e. subject, predicate , and object. We applied topFiberM to Semantic Web Dog Food\footnote{https://old.datahub.io/dataset/semantic-web-dog-food} (SWDF) dataset and also Asso. 
 The dataset is about conferences and papers related to Semantic Web during years 2001 to 2015. We presented the graph as a set of frontal slices (i.e. each slice represents one predicate). 
 For Asso it is required to remove the columns which are all-zeros other wise it will not be able to finish. We considered the predicates that are represented 1000 times or more and the rank used was 100. 
 The running times show that topFiberM is faster than Asso with two order of magnitudes ( in average 128 times) and faster than GreConD on average 838 while giving slight improvement in coverage (c=0.4535 for \grecond, 0.4460 for Asso, and 0.4538 for topFiberM). \grecond originally implemented MATLAB we converted that code to R.
 We tried \grecondplus MATLAB implementation with critical parts implemented in C but it turned out to be much slower than \grecond (It took more than 20 hours in processing only the first two matrices).
\begin{figure}[!ht]
    \centering
    \includegraphics[width=\columnwidth]{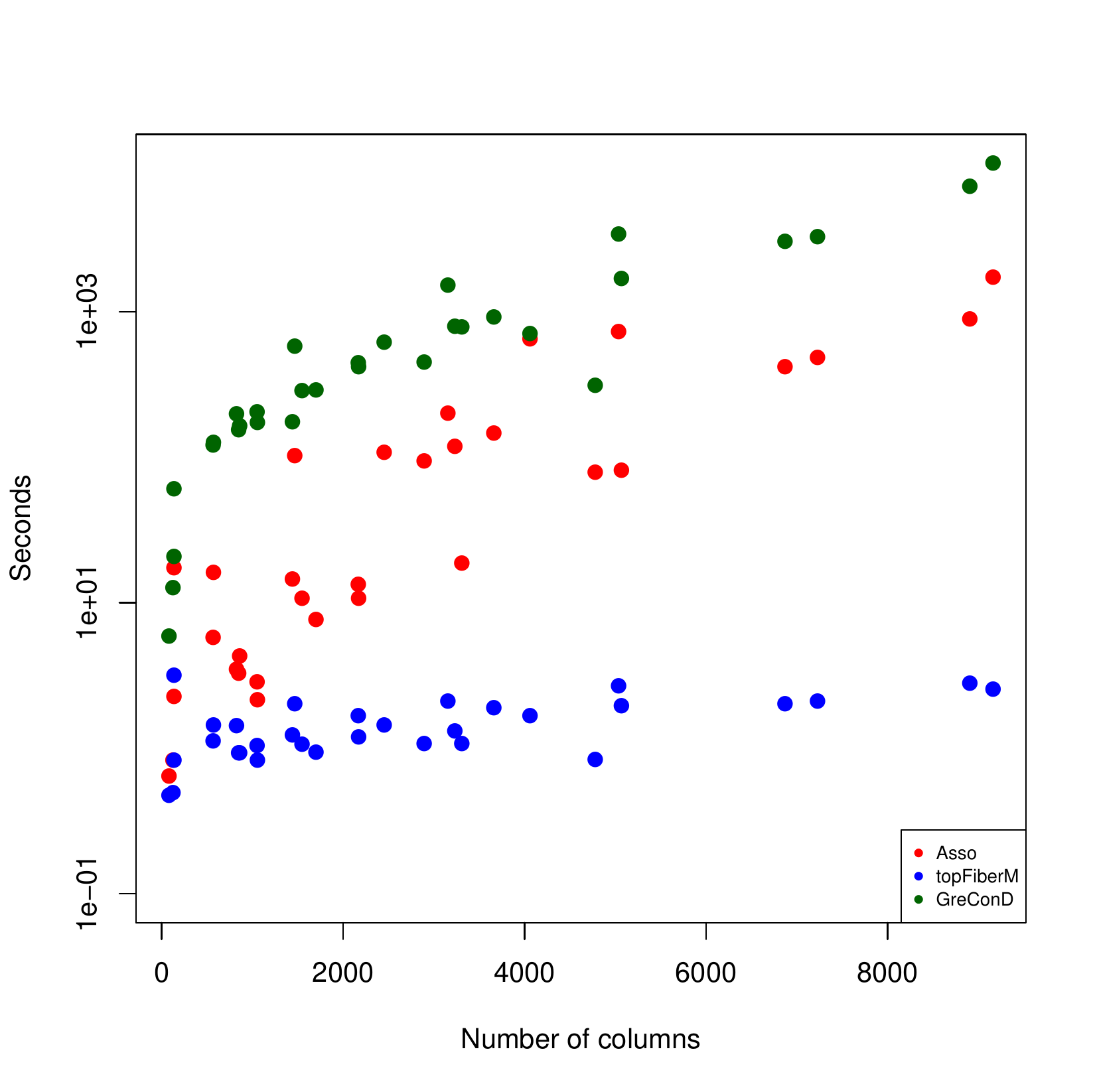}
    \caption{Run times on seconds for applying Asso, topFiberM, and \grecond to SWDF dataset predicates.
    }
    \label{fig:cmpAsso_topFiber}
\end{figure}

 The running times for the 31 predicates are shown in Figure \ref{fig:cmpAsso_topFiber}. We used a standard core i5 laptop of 16GB RAM running Windows 10 for the experiment. 
From this figure it is clear that topFiberM is scalable in terms of number of columns. The predicate the take the longest time by topFiberM is \texttt{rdf:type} predicate which contains the maximum number of nonzeros.

\section{Conclusion and Future Work}
We proposed topFiberM a new BMF algorithm. 
topFiberM finds factors in rows or columns in a greedy way then extends them to rectangles if possible.
Our algorithm showed better performance in AFP and DBP. topFiberM is much faster than the famous \emph{Asso} algorithm.
Future work include applications of our algorithm in RDF graphs and extending it to Boolean Tensor Factorization.

\section{Acknowledgements}
We thank Martin Trnecka for providing the datasets and MATLAB code of \grecond and \grecondplus.
\bibliographystyle{splncs04}
\bibliography{topFiberM}

\end{document}